\documentclass[aps,prb,amsmath]{revtex4}

\usepackage{graphicx}
\usepackage{bm}

\begin{document}

\title{Wetting effect and morphological stability in growth of
short-period strained multilayers}
\author{Zhi-Feng Huang}
 \email{huang@csit.fsu.edu}
 \altaffiliation[Present address: ]{School of Computational Science
and Information Technology, Florida State University, Tallahassee,
Florida 32306-4120}
\author{Daniel Kandel}
 \email{daniel.kandel@weizmann.ac.il}
\altaffiliation[Permanent address: ]{Department of Physics of 
Complex Systems, Weizmann Institute of Science, Rehovot 76100, Israel}
\author{Rashmi C. Desai}
 \email{desai@physics.utoronto.ca}
\affiliation{%
Department of Physics, University of Toronto,
Toronto, Ontario, Canada M5S 1A7
}%
\date{Submitted to Appl. Phys. Lett. 24 January 2003; accepted 1 May 2003}
\begin{abstract}

We explore the morphological stability during the growth of strained
multilayer structures in a dynamical model which describes the coupling
of elastic fields, wetting effect, and deposition process. We
quantitatively show the significant influence of the wetting effect on
the stability properties, in particular for short-period multilayers.
Our results are qualitatively similar to recent experimental observations
in AlAs/InAs/InP(001) system. We also give predictions for
strain-balanced multilayers.

\end{abstract}

\maketitle

Strained periodic multilayer films composed of different material
layers have attracted much attention since such material
structures can have tunable electronic properties. Among the actively
investigated systems is the multilayer with alternating
tensile/compressive layers coherently grown on a substrate, resulting
in a modulated structure like short-period superlattice [e.g.,
GaAs/InAs \cite{cheng92} or AlAs/InAs
\cite{mirecki97,norman00,norman02} on InP(001)] or a multiple quantum
well [e.g., GaInP/InAsP on InP(001)] \cite{ponchet94}. Generally
morphological instability occurs in such coherent lattice-mismatch
multilayer structures, driven by a gradual release of misfit-generated
stresses when dislocations are absent. Consequently, undulations are
observed in the strained layers, instead of ideally flat interfaces.
This stress-driven morphological instability can lead to {\it lateral}
compositional modulations in {\it vertical} short-period superlattices,
\cite{cheng92,mirecki97,norman00,norman02} as a result of layer
thickness modulations (caused by interface rippling) and different
material components in adjacent layers. \cite{ponchet94,norman00} The
spontaneous formation of these lateral modulations is also a promising
way to self-organize fabrication of low-dimensional quantum
heterostructures, especially quantum wires.
\cite{cheng92,mirecki97,norman00}

Although understanding of the detailed growth mechanism in multilayer
structures is important and of much interest for both fundamental
studies and device applications, it is still far from complete due to
the complexity of the system. This complexity arises from the coupling
of strain fields in different layers, and the nonequilibrium nature of
the growing film for which material deposition rates play an important
role in the pattern formation. Previous theoretical analyses use some
approximations to determine the elastic fields, treating each buried
island \cite{tersoff96} or nonplanar interface \cite{shilkrot00} of the
multilayer as a misfitting inclusion in a semi-infinite homogeneous
medium. Recently, a general method has been developed to directly
calculate the elastic state of the multilayer system. \cite{huang02}

Wetting effects have been considered in the context of single-layer
strained film growth, \cite{tersoff91,kandel00} but not for
multilayers. In this letter, we attempt to remedy this. The wetting
effect arises from the change of material properties across interlayer
interfaces within the system, and is especially important for thin
composite layers. Here, we incorporate the wetting
effect (arising from nonlinear elastic contributions \cite{kandel00})
in the early morphological evolution of growing multilayer films. We
focus on the stability properties of the system and on comparison with
recent experiments.

We consider a three-dimensional coherent multilayer structure with $k$
stacking layers deposited on a semi-infinite substrate. In a general
case, each composite layer $i$ ($1 \leq i < k$) could consist of a
material different from all the other layers and substrate, with
average thickness $l_i$ and height profile $z=h_i(x,y)=\sum_{j=1}^{i}
l_j + \sum_{\bf q} \hat{h}_i({\bf q}) \exp(iq_x x+iq_y y)$. The
film/substrate interface is described by $\hat{h}_0({\bf q})$, and the
top strained layer $k$ by a surface morphological perturbation
$\hat{h}_k({\bf q},t)$ and by an average thickness $l_k=v_k t$ where
$v_k$ is the deposition rate. 

The basic state of the system is chosen to be a uniform film of flat
growing surface and planar underlying interfaces, with all strains zero
except for $\bar{u}_{zz,i}=\bar{u}_i=\epsilon_i (1+\nu_i)/(1-\nu_i)$
due to Poisson relaxation in the $z$ direction. $\epsilon_i$ and
$\nu_i$ are the misfit and Poisson ratio of layer $i$, respectively.
The elastic free energy of the multilayer is written as ${\cal F}_{\rm
el}=\int_{-\infty}^{h_0} d^3r {\cal E}_0 + \sum_{i=1}^{k}
\int_{h_{i-1}}^{h_i} d^3r {\cal E}_i$, where the elastic energy density
for each layer $i$ can be generally expressed as ${\cal E}_i=\bar{\cal
E}_i + \delta {\cal E}_i$. Here $\bar{\cal E}_i$ is the energy density
of the basic state and $\delta {\cal E}_i$ is the change in energy
density due to a perturbation. Each of the two terms includes a linear
contribution, obtained from linear elasticity, and a nonlinear
contribution. The latter is often ignored, but is crucial for the
wetting effect. \cite{kandel00} It leads to the dependence of
$\bar{\cal E}_i$ on layer thickness $h_i(x,y)-h_{i-1}(x,y)$. This
dependence is attributed to the coupling between the elastic relaxation
of layer $i$ and that of the underlying layer or substrate $i-1$. It
was shown \cite{kandel00} that the thickness dependence of $\bar{\cal
E}_i$ enhances the wetting of the substrate by the film in the case of
a single layer, and consequently stabilizes the flat surface.

For the energy density of the perturbed state, $\delta {\cal E}_i$, we
only keep the linear elasticity contribution $\delta {\cal E}_i^{\rm
lin}=\bar{\sigma}_{\alpha\beta,i} \delta u_{\alpha\beta,i} +
\lambda_{\alpha\beta\xi\rho} \delta u_{\alpha\beta,i}\delta
u_{\xi\rho,i}/2$ (with $\sigma_{\alpha\beta,i}$ the stress tensor 
and $\lambda_{\alpha\beta\xi\rho}$ the elastic
modulus tensor), since the nonlinear contribution is of higher order in
the perturbed strains $\delta u_{\alpha\beta,i}$. Thus, we have ${\cal
E}_i=\bar{\cal E}_i + \delta {\cal E}_i^{\rm lin}$. Using the results
of Ref. \onlinecite{huang02} for the first-order surface elastic energy
density, $\delta \hat{\cal E}_k^{\rm lin}$, we obtain the first-order
evolution equation for $\hat{h}_k({\bf q},t)$:
\begin{eqnarray}
\partial \hat{h}_k / \partial t &=& \left \{ \sigma_k -
\Gamma_k q^2 [d^2 f_{{\rm el},k}^{(0)}/dh_k^2]
_{h_k=\bar{h}_k(t)} \right \} \hat{h}_k \nonumber\\
&-& \Gamma_k E' \epsilon_k q^3 e^{-qv_k t}
\left [ \sum\limits_{j=1}^{k-1} \left ( \epsilon_{j+1}
-\epsilon_j \right ) e^{-q\sum\limits_{i=j+1}^{k-1} l_i}
\hat{h}_j + \epsilon_1 e^{-q\sum\limits_{i=1}^{k-1} l_i}
\hat{h}_0 \right ]
\label{eq-h}
\end{eqnarray}
with $\sigma_k=\Gamma_k \left [ E'\epsilon_k^2 q^3 - \tilde{
\gamma}_k(0) q^4 \right ]$ and $f_{{\rm el},i}^{(0)}
(h_{i-1},h_i)=\int_{h_{i-1}}^{h_i} dz \bar{\cal E}_i(h_i-h_{i-1},z)$.
The derivatives of the latter quantity with respect to $h_i$ are taken
with fixed $h_{i-1}$. Here we assumed that surface diffusion is the 
dominant mass transport process, with negligible bulk and interlayer 
diffusion as well as frozen buried interfaces. \cite{huang02} We used a
step-function form for surface stiffness ($\tilde{\gamma}$) and misfit
stresses ($\bar{\sigma}_{xx,i}=\bar{\sigma}_{yy,i}$) at the interlayer
interfaces. \cite{kandel00} In Eq. (\ref{eq-h}) $\Gamma_k$ is the
surface mobility and $\tilde{\gamma}_k(0)$ is the surface stiffness at
the orientation of the unperturbed surface. The average surface
position is $\bar{h}_k=\sum_{i=1}^{k-1}l_i + v_k t$, and
$E'=2E(1+\nu)/(1-\nu)$ with Young's modulus $E$ and Poisson ratio $\nu$
assumed to be identical for all the layers and substrate. Note that the
corresponding results for single-layer film can be recovered from Eq.
(\ref{eq-h}) with $k=1$.

The evolution Eq. (\ref{eq-h}) is similar to that without the
wetting effect, \cite{huang02} showing explicitly the influence of
interface morphologies and elastic properties of different buried
layers on the top surface profile. The important results of
this work arise from the term $d^2 f_{{\rm el},k}^{(0)}/dh_k^2$,
which represents the wetting effect. The exact solution of Eq.
(\ref{eq-h}) has a form similar to the solution in the absence of the
wetting effect. \cite{huang02} The surface profile $\hat{h}_k$ can be
expressed in terms of the product of $(2 \times 2)$ characteristic
matrices $\bar{\bf{L}}_j$ ($3 \leq j \leq k$) as well as the profiles
$\hat{h}_1$ and $\hat{h}_2$ of first and second deposited layers. The
elements of the bottom row of $\bar{\bf{L}}_j$ are $1$ and $0$ and
those of the top row are $\bar{L}_{j,1}(q)$ and $\bar{L}_{j,2}(q)$,
where
\begin{eqnarray}
\bar{L}_{j,1}(q) &=& \mu_j(q,l_j/v_j) \left [ 1-\Gamma_j E' q^3
\epsilon_j (\epsilon_j - \epsilon_{j-1}) G_j(q,l_j/v_j) \right ]
-\bar{L}_{j,2}(q)/\mu_{j-1}(q,l_{j-1}/v_{j-1}), \nonumber\\
\bar{L}_{j,2}(q) &=& -\mu_j(q,l_j/v_j) e^{-q l_{j-1}}
[\Gamma_j \epsilon_j G_j(q,l_j/v_j)] / [\Gamma_{j-1}
\epsilon_{j-1} G_{j-1}(q,l_{j-1}/v_{j-1})].
\label{eq-L}
\end{eqnarray}
Here
\begin{eqnarray}
\mu_j(q,l_j/v_j) &=& e^ {\sigma_j l_j/v_j - \Gamma_j q^2
\int_0^{l_j/v_j} dt' [d^2 f_{{\rm el},j}^{(0)}/dh_j^2]_
{l_j=v_jt'} }, \label{eq-mu} \\
G_j(q,l_j/v_j) &=& \int_0^{l_j/v_j} dt'
e^ { -(qv_j+\sigma_j)t' + \Gamma_j q^2 \int_0^{t'} dt''
[d^2 f_{{\rm el},j}^{(0)}/dh_j^2]_{l_j=v_jt''} }.
\label{eq-G}
\end{eqnarray}

In the following these general results, which can apply to both
periodic and nonperiodic systems, are used to determine the
morphological stability of periodic tensile/compressive multilayer
structures, and in particular short-period superlattices.
\cite{cheng92,mirecki97,norman00,norman02} In this case the multilayer
film is composed of alternating $A$ and $B$ layers, with their
associated parameters such as nonzero misfits $\epsilon^*_{A(B)}$,
layer thicknesses $l^*_{A(B)}$, surface mobilities $\Gamma^*_{A(B)}$,
deposition rates $v^*_{A(B)}$ and surface stiffnesses
$\gamma^*_{A(B)}$. Here we have expressed all the parameters in
dimensionless form: $l_A^0 = \tilde{\gamma}_A(0)/(E'\epsilon_A^2)$ and
$v_A^0=\Gamma_A E'^3 \epsilon_A^6 / \tilde{\gamma}_A^2(0)$, as well as
$\epsilon^*_{B} =\epsilon_B/\epsilon_A$, $\Gamma_B^*=\Gamma_B/\Gamma_A$
and $\gamma_B^*=\tilde{\gamma}_B(0) / \tilde{\gamma}_A(0)$. To evaluate
the wetting effect, we assume the form $d f_{{\rm el},j} ^{(0)}/dh_j =
\bar{\cal E}_j^{\rm lin} \{ 1-\chi_j \exp [-(h_j-h_{j-1})/h_{{\rm
ML},j}] \}$ (for $h_j>h_{j-1}$) appropriate for the single-layer case,
(according to the qualitative modeling calculations in ref.
\onlinecite{kandel00}, which yield this general functional form of 
wetting elastic energy for various lattice structures with diagonal
bonds), where $\bar{\cal E}_j^{\rm lin}=E\epsilon_j^2/(1-\nu)$
is the linear elastic energy density of the basic state and $h_{{\rm
ML},j}$ is the 1 ML thickness of layer $j$ ($A$ or $B$). We also assume
$\chi_j= \chi=0.05$ for all strained layers.

For this periodic $A/B$ system, the characteristic matrices $\bar{\bf
L}_j$ presented in Eq. (\ref{eq-L}) are reduced to two types, $\bar{\bf
L}_{A(B)}$, and the effective stability (for large number of grown
layers $k \gg 1$) is determined by the maximal magnitude of the
eigenvalues $\lambda$ of the matrix $\bar{\bf L}_B\bar{\bf L}_A$, as
shown in Ref. \onlinecite{huang02}.  When $\max(|\lambda|)>1$, the
initial perturbations are amplified after further deposition, leading
to a morphological instability; otherwise, the perturbations decay
during the growth, leading to a stable flat film profile.

\begin{figure}
\resizebox{6.2cm}{!}{\includegraphics{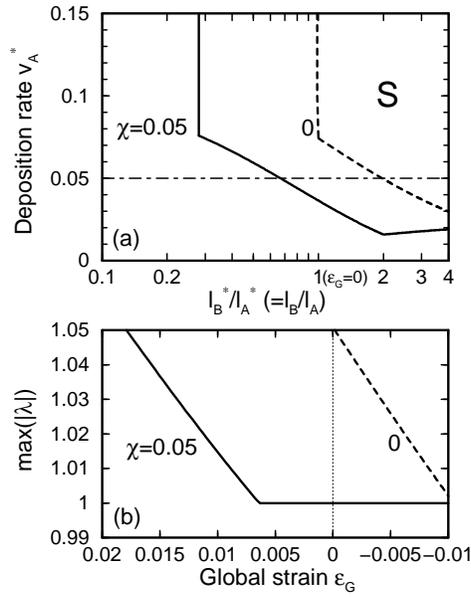}}
\caption{\label{fig-lambda}(a) Stability diagram of $l_B^*/l_A^*$
vs rescaled deposition rate $v_A^*$, with ($\chi=0.05$, solid
line) or without ($\chi=0$, dashed line) the wetting effect. The
parameters are analogous to those of AlAs/InAs on InP(001)
superlattices growth: $\Gamma_B^*=0.001$, $l_A^*+l_B^*=0.3$,
$\epsilon_B^*=-1$, $v_B^*/v_A^*=1$, and $\gamma_B^*=1$.
(b) The maximum value of $|\lambda|$ as a function of multilayer
global strain $\epsilon_G$, with the same parameters of (a) except
for a fixed deposition rate $v_A^*=0.05$, corresponding to the
horizontal dot-dashed line in (a).}
\end{figure}

Consequently, the stability diagrams for various growth and material
parameters can be determined, as presented in Fig. \ref{fig-lambda},
where in order to compare with experiments, the parameters are chosen
to qualitatively represent those of the growth of AlAs/InAs/InP(001) short-period
superlattices. \cite{norman00,norman02} Note that the global strain of
the whole multilayer film can be defined as $\epsilon_G=\epsilon_A
[1+(l_B/l_A)\epsilon_B^*] / (1+l_B/l_A)$. Thus for $\epsilon_B^*=-1$
(as in Fig. \ref{fig-lambda}), $l_B/l_A=1$ corresponds to
$\epsilon_G=0$, while $l_B/l_A>1$ or $<1$ to tensile or compressive
films. Similar to previous calculations without the wetting effect,
\cite{huang02} the stability diagram shows a clear asymmetry: the
multilayer film is more {\it stable} when the global strain
$\epsilon_G$ has the same sign as the misfit of the layers with {\it
smaller} surface mobility. For the system of AlAs/InAs/InP(001),
$\Gamma_{\rm InAs} (\Gamma_A) \gg \Gamma_{\rm AlAs}(\Gamma_B)$, and
then the globally tensile film ($\epsilon_G<0$) is more stable since
$\epsilon_ {\rm AlAs}<0$, as seen in Fig. \ref{fig-lambda}(a) and in
experiments. Furthermore, as shown in Fig. \ref{fig-lambda}(a),
without the wetting effect (dashed line), the flat film is stable only
when the global strain is tensile. The situation is qualitatively
different when the wetting effect is included [solid line in Fig.
\ref{fig-lambda}(a)]: both the globally tensile and compressive flat
films can be stabilized, although the asymmetry in the stability
diagram is still obvious.

The earlier result with the consideration of the wetting effect is
qualitatively similar to experimental observations \cite{norman00,norman02}
in the following way: We plot [Fig. \ref{fig-lambda}(b)]
$\max(|\lambda|)$ as a function of global strain, $\epsilon_G$, for
$v_A^*=0.05$, which corresponds to the horizontal dot-dashed line in
Fig. \ref{fig-lambda}(a). To evaluate $\epsilon_G$, we set
$\epsilon_A=3.2\%$ according to the InAs/InP misfit. As explained
earlier, when the maximal value of $\lambda$ is equal to $1$, the
multilayer film is stable, with a smooth surface; when it is larger
than $1$, the film is unstable, leading to a rough surface. Also, the larger
the value of $\max(|\lambda|)$, the rougher the surface that results,
as obtained qualitatively from our theoretical results 
\cite{huang02} $\hat{h}_k \propto |\lambda|^{k/2}$ for $k \gg 1$.
Thus, this figure can be compared with the recent rms roughness
measurement of Norman {\it et al.}, \cite{norman02} where the surface
is flat or with small roughness for samples under global tension or
small compression, but shows a rapidly increasing degree of roughness for
larger compression. This experimental finding can be qualitatively inferred
from our result including wetting effect [solid line in Fig.
\ref{fig-lambda}(b)], but not from the result without wetting (dashed
line).

We also find that with the wetting effect, the stability boundary value
of $l_B/l_A$, above which stabilization is possible, depends
sensitively on $l_A^*+l_B^*$, but only very slightly on $\Gamma_B^*$.
The stability boundary value is closer to $1$ (the value in the absence
of the wetting effect) when $l_A^*+l_B^*$ is larger, showing that the
wetting effect has less influence for thicker layer periods.

\begin{figure}
\resizebox{6.2cm}{!}{\includegraphics{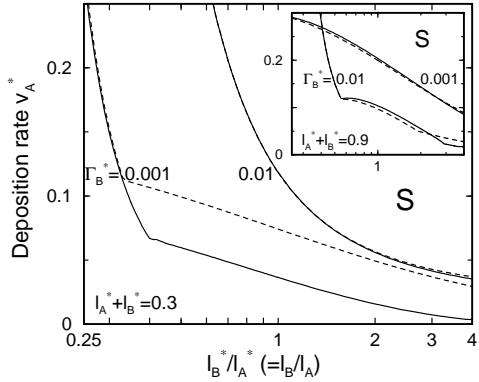}}
\caption{\label{fig-v-lBA-bal}Stability diagram of $l_B^*/l_A^*$
vs $v_A^*$ in the strain-balanced condition, with ($\chi=0.05$,
solid line) or without ($\chi=0$, dashed line) the wetting effect.
The effects of different relative surface mobilities ($\Gamma_B^*
=0.01$ and $0.001$) and multilayer periods [$l_A^*+l_B^*=0.3$ and
$0.9$ (inset)] are shown. Other parameters are $v_B^*/v_A^*=1$ and
$\gamma_B^*=1$ as in Fig. \ref{fig-lambda}.}
\end{figure}

Since some multilayer systems are grown under strain-balance conditions
($\epsilon_G=0$), \cite{ponchet94} it is interesting to study the
corresponding stability properties. The diagrams of $l_B^*/l_A^*$
($=-1/\epsilon_B^*$ for $\epsilon_G=0$) versus $v_A^*$ are shown in
Fig. \ref{fig-v-lBA-bal}, where the stability asymmetry remains and is
independent of relative mobility $\Gamma_B^*$, indicating more
stability for asymmetric structures. \cite{huang02} From Fig.
\ref{fig-v-lBA-bal}, one can see that with the strain-balance
constraint, there is no qualitative change due to the wetting effect.
Also, for short period (e.g., $l_A^*+l_B^*=0.3$) and large mobility
difference (e.g., $\Gamma_B^*=0.001$), the wetting effect tends to
render the film more stable.

In summary, we have studied the interplay of elasticity, wetting layer
effects and material deposition in coherent multilayer growth, and
determined the stability and morphological evolution of growing films.
The important influence of the wetting effect for short-period
multilayers has been calculated, reproducing recent experimental
measurements. We hope these results will be helpful for practical
growth of low-dimensional nanostructures.

\newpage

\end{document}